\documentclass{article}
\usepackage[numbers]{natbib}
\usepackage {arxiv}

\usepackage{cite}
\usepackage{stfloats}  
\usepackage{floatrow}
\usepackage{amsmath,amssymb,amsfonts}
\usepackage{algorithm}
\usepackage{algorithmic}
\usepackage{textcomp}
\usepackage{xcolor}
\usepackage{float}
\usepackage{booktabs}
\usepackage{enumitem}
\usepackage{setspace}
\usepackage[]{graphicx}
\usepackage{amsmath}    
\usepackage{amssymb}
\usepackage{epsfig}
\usepackage{epstopdf}
\usepackage{subfigure}
\usepackage{comment}
\usepackage{tabularx}
\usepackage{threeparttable}
\usepackage{adjustbox}
\usepackage{soul}
\usepackage{url}
\usepackage{hyperref} 
\usepackage{graphicx}

\title{\LARGE \bf
Predicting Early and Complete Drug Release from Long-Acting Injectables Using Explainable Machine Learning}

\author{Karla N. Robles\\
TIGER Institute of Advanced Materials\\
Tennessee State University\\
Nashville, TN, USA \\
\and 
{\bf Manar D. Samad}\\
Department of Computer Science\\
Tennessee State University\\
Nashville, TN, USA\\
 }

\begin{document}


\maketitle

\begin{abstract}

Polymer-based long-acting injectables (LAIs) have transformed the treatment of chronic diseases by enabling controlled drug delivery, thus reducing dosing frequency and extending therapeutic duration. Achieving controlled drug release from LAIs requires extensive optimization of the complex underlying physicochemical properties. Machine learning (ML) can accelerate LAI development by modeling the complex relationships between LAI properties and drug release. However, recent ML studies have provided limited information on key properties that modulate drug release, due to the lack of custom modeling and analysis tailored to LAI data. This paper presents a novel data transformation and explainable ML approach to synthesize actionable information from 321 LAI formulations by predicting early drug release at 24, 48, and 72 hours, classification of release profile types, and prediction of complete release profiles. These three experiments investigate the contribution and control of LAI material characteristics in early and complete drug release profiles. A strong correlation ($>$0.65) is observed between the true and predicted drug release in 72 hours, while a 0.87 F1-score is obtained in classifying release profile types. A time-independent ML framework predicts delayed biphasic and triphasic curves with better performance than current time-dependent approaches. Shapley additive explanations reveal the relative influence of material characteristics during early and for complete release, which fill several gaps in previous \emph{in-vitro} and ML-based studies. The novel approach and findings can provide a quantitative strategy and recommendations for scientists to optimize the drug-release dynamics of LAI. The source code for the model implementation is publicly available in \footnote{\url{https://github.com/mdsamad001/Drug_Release_Dynamics_Prediction/}.}  


\end{abstract}

\section{Introduction} \label{introduction}

Traditional research and development (R\&D) efforts in industry and manufacturing are often impeded by high costs, low success rates, and prolonged research cycles \citep{xing2020industrialAI}. Compared to traditional methodologies carried out in laboratories, R\&D based on artificial intelligence (AI) \&D can better optimize product design and support accelerated material selection and function tests while reducing time and cost~\citep{xing2020industrialAI}. In materials science, machine learning (ML), a subset of AI tools, can accelerate innovation by predicting the behavior and characteristics of materials based on their composition\citep{adetunla2024AIinMaterials}. Therefore, the characteristics of the target material can be rapidly prototyped using knowledge from limited laboratory data and advanced ML~\citep{adetunla2024AIinMaterials}. A remarkable area of materials science has led to next-generation drug delivery technologies~\citep{elTanani2025revolutionizing}. Among these, long-acting polymer-based drug carriers improve therapeutic outcomes by reducing off-target toxicity while increasing the bioavailability of small molecules and biological drugs~\citep{sultana2022nano}. In this context, long-acting polymer-based injectable therapeutics require clinical approval for administration to treat chronic conditions such as neurological disorders, diabetes, and cancer \citep{obrien2021challenges, Witika2025}.

Before clinical approval, the development of novel long-acting polymeric drug carriers involves complex optimization of numerous formulation parameters \citep{park2019injectable}, including particle size and drug encapsulation efficiency. The amount of drug released over time, known as the drug release profile, is a critical property of drug carriers. An optimized drug release profile is designed to maintain drug concentrations within a therapeutic window, the range between therapeutic and toxicity thresholds ~\citep{yoo2020burstrelease}. For example, a burst release of high concentration of drug~\citep{park2019injectable} can exceed toxicity thresholds, resulting in unwanted side effects~\citep{yoo2020burstrelease}. However, the burst release phenomenon is difficult study~\citep{park2019injectable} because the definition of burst varies across drug types and is usually revealed only by \textit{in-vitro} characterization. Furthermore, drug release profiling is the longest part of polymeric drug carrier characterization, which can take weeks to months. The variability in drug release duration, inconsistent sampling time, and heterogeneity in drug release requirements make it challenging to analyze the contributions of material characteristics to drug release profiles. Recent ML studies have used traditional methods and revealed a strong time-dependence in release profiles, thus overshadowing the relative contributions of material characteristics~\citep{Bannigan2023}. Likewise, burst release is defined using a drug-specific criterion based on the magnitude of release at a given time. Instead, we hypothesize that the drug release profiles of LAI, unlike short and burst release, can take varying drug release curves as a result of the influence of material characteristics. The time-independent curve of the release profile can reveal the contributions of LAI material characteristics, which were previously obscured by time-dependent modeling. This paper presents a novel time-independent data transformation and a custom ML model to investigate the relationship between material characteristics and LAI drug-release profiles.

 
The organization of this paper is as follows. Section \ref{background} provides a summary of current ML methods used to predict drug release. Section \ref{methods} outlines the proposed models and the experiments conducted in this study. Section \ref{results} presents and compares the experimental results. Finally, Section \ref{discussion} summarizes the key findings, discusses insights derived from the results, suggests possible directions for future research, and concludes the article.

\section{Background} \label{background}

Long-acting injectables (LAI) improve the treatment of chronic diseases through sustained drug delivery, thereby reducing the frequency of drug administrations and the burden of treatment for patients \citep{baryakova2023overcoming}. LAIs are developed to meet the unique therapeutic requirements of a target disease, including specific rates and concentrations of drug delivery. However, \textit{in-vitro} LAI optimization entails iterative, costly, and time-intensive characterization of drug release rates because identifying the key drivers of drug release remains a challenge \citep{Park2025PLGA}. 

In this context, ML offers powerful computational solutions to modeling such multidimensional problems. Recent literature reviews have extensively examined the application of ML in biomaterial research~\citep{Meyer2022}, including the study of polymer-based microparticles for drug delivery \citep{Bao2025} and drug release \citep{Aghajanpour2025}. Some applications of traditional ML predict microparticle characteristics such as particle size and drug loading~\citep{alqarni2025predicting, noorain2023mlplga, Rezvantalab2024}. Several groups have developed methods to predict partial \citep{Husseini2025} and time-dependent \citep{Bannigan2023}  drug release profiles. However, a complete prediction of the drug release profile has not been well explored due to the lack of high-quality data and robust methodological approaches. A data-driven model that learns the relationship between material characteristics and drug release can be a valuable LAI development tool, substantially reducing the time and resources required for optimization.  

This paper expands on recent successes \citep{Deng2023, Bannigan2023} of ML in modeling complex non-linear relationships between biomaterial properties and drug release dynamics. Among ML methods for predicting drug release profiles, decision tree-based methods are particularly successful~\citep{deAzevedo2017, Bannigan2023, Zawbaa2016}. For example, Bannigan et al.~\citep{Bannigan2023} compared 11 traditional ML methods to predict drug release profiles, in which a light gradient-boosted tree model (GBDT) achieved the best performance. A key limitation of this and similar studies is their reliance on initial \citep{Bannigan2023} or prior \citep{Husseini2025} drug release information to forecast the subsequent drug release profile. The requirement of preceding experimental data in time and drug-specific criterion for release time are some impediments to developing robust models for discovering key material characteristics in this process.

Few studies have explored the application of vanilla deep learning to model drug release profiles \citep{Zawbaa2016, Salma2021, Husseini2025, Deng2023} without customizing for data-specific requirements. Collectively, these approaches have investigated the performance of various model architectures, including MLPs (68 formulations) \citep{Zawbaa2016}, DNNs \citep{Husseini2025}, CNNs, LSTMs, and GANs \citep{Salma2021}. Deng et al. \citep{Deng2023} have used complex deep learning models with 10 hidden layers and the 'black-box' ResNet architecture with 20 hidden layers to predict drug release for polymer-based microparticles (283 formulations). There are two major limitations of deep learning models. First, the sample size is often insufficient for deep learning to overcome overfitting. Second, the results of black-box deep models are challenging to interpret, but explainability is necessary to provide deeper insights into the prediction of drug release dynamics. A common approach in traditional and deep learning studies \citep{Zawbaa2016, Salma2021, Husseini2025, Deng2023, Bannigan2023, Zhang2025} is to use time values as input variables along with static material characteristics. Drug release certainly increases with time. However, the dependence on time of exisiting methods is so strong that they would not maintain predictive performance using only material characteristics. Therefore, we argue that strong influence of time in predicting the drug release profile, as presented in the SHAP feature analysis~\citep{Bannigan2023}, overshadows the contributions of material characteristics to shaping the drug release profile. 

\begin{figure}[t]
    \centering
    \includegraphics[width=1.0\linewidth]{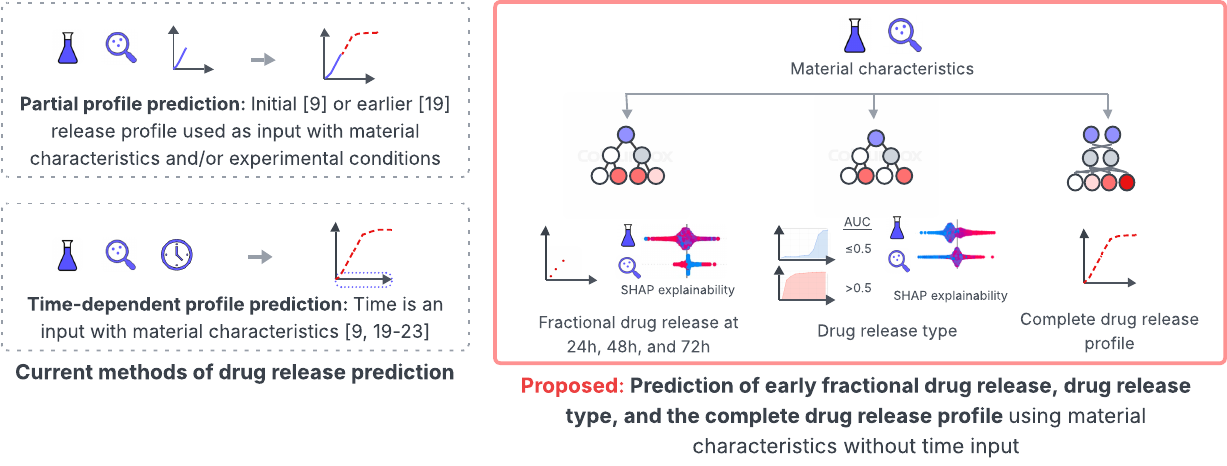}
    \caption{Overview of standard approaches to predicting drug release and methodological contributions of this article.}
    \label{prediction_methods}
\end{figure}

This paper proposes a novel data transformation and custom ML framework to predict the different curves of LAIs release profiles using material characteristics. The proposed framework addresses several key limitations of existing work, as summarized in Figure ~\ref{prediction_methods} to make the following contributions.  


\begin{itemize}
    \item     We build our models from multi-experiment and multi-source data \citep{bao2025datasetPLGAMPs} to improve the reproducibility and generalizability of drug release prediction. 
       \item Fractional drug release is predicted at 24, 48, and 72-hour time points to investigate the influence of material characteristics during the early release period when burst release possibly takes place.
    \item  A novel area under the curve metric is proposed to characterize the time-independent curve of drug release. LAI drug release profiles are categorized into two distinct classes to investigate the effects of material characteristics in delaying drug release. 
 \item The curve of the complete drug release profile is predicted without preceding release information or requiring time as an input.

    \item  We have customized a recurrent neural network model to predict the complete curve of the drug release profile from the collective contributions of material characteristics.
    \item We present SHAP analyses to explain the influence of individual material characteristics for each predictive model. The relative influence of material characteristics on shaping the drug release profiles of polymer-based LAIs are presented and compared with current \emph{in-vitro} and ML studies.

\end{itemize}

\section{Methods} \label{methods}

This section provides a detailed description of the data set, data preparation, experiments, and evaluation of the models to predict drug release from LAI material characteristics.

\subsection{Data set and data preparation}
The data set of drug-loaded PLGA microparticles has recently been compiled from 113 publications using keywords such as PLGA microparticles, microspheres, and drug delivery~\citep{bao2025datasetPLGAMPs}. The data set includes 321 drug release profiles of 89 drugs along with the corresponding formulation parameters and microparticle characteristics. These formulation parameters and microparticle characteristics are listed in Table~\ref{features} and are used as input features to predict drug release.

\begin{table}[t]
\centering
\caption{Summary of formulation parameters and microparticle characteristics in the data set, adapted from \citep{bao2025datasetPLGAMPs}}.
\label{features}
\begin{tabular}{ll}
\toprule
\textbf{Feature} & \textbf{Description}\\ 
\midrule
Formulation Method & Emulsion method used in microparticle preparation \\
Drug MW & Drug molecular weight\\
Drug TPSA & Drug topological polar surface area \\
Drug LogP & Drug log partition coefficient between aqueous and lipophilic phases  \\
Polymer MW & Polymer molecular weight of the carrier\\
LA/GA & Polymer molar ratio of lactide to glycolide \\
Initial DPR & Initial drug-to-polymer weight ratio in microparticle formulation \\
Particle size & Diameter of microparticles \\
Encapsulation Efficiency & Percent of initial drug encapsulated in microparticles \\
Loading Capacity & Percent of drug-to-microparticle mass \\
Solubility Enhancer Conc. & Solubility enhancer concentration in \textit{in-vitro} media \\
\bottomrule
\end{tabular}
\end{table}

The data set includes drug release profiles of a wide range of durations (min: 72 hours, max: 238 days, mean: 30 $\pm$ 25 days), which are sampled at various time points. All release profiles have achieved at least 60\% of the maximum drug release.
To ensure data uniformity in predictive modeling, drug release profiles are normalized and interpolated over time to produce distinct, comparable curves as follows. The duration of each drug release is scaled between 0 and 1 using Min-Max normalization, where each normalized time point ($t_{norm}$) is defined below.
\begin{equation}
t_{\text{norm}} = \frac{t - t_{\min}}{t_{\max} - t_{\min}}
\end{equation}
Here, $t$ is the original time point, and $t_{min}$ and $t_{max}$ are the minimum and maximum time points of the drug release profile, respectively. Linear interpolation is applied to the drug release profiles to ensure uniformly sampled and fixed length input to the predictive models. Given two known release values $y_{0}$ and $y_{1}$ at time points $t_0$ and $t_1$, respectively, a release value at a time point $t_0 < t_{int} < t_1$ can be interpolated as follows.
\begin{equation}
y_{int} = y_{0} + \frac{t_{int} - t_{0}}{t_1 - t_0} \cdot (y_{1} - y_{0})
\end{equation}
Thus, time interpolation maintains the rate of release at a given point to complete the curve of drug release profile.

\begin{figure}[t]
    \centering
    \includegraphics[width=\linewidth]{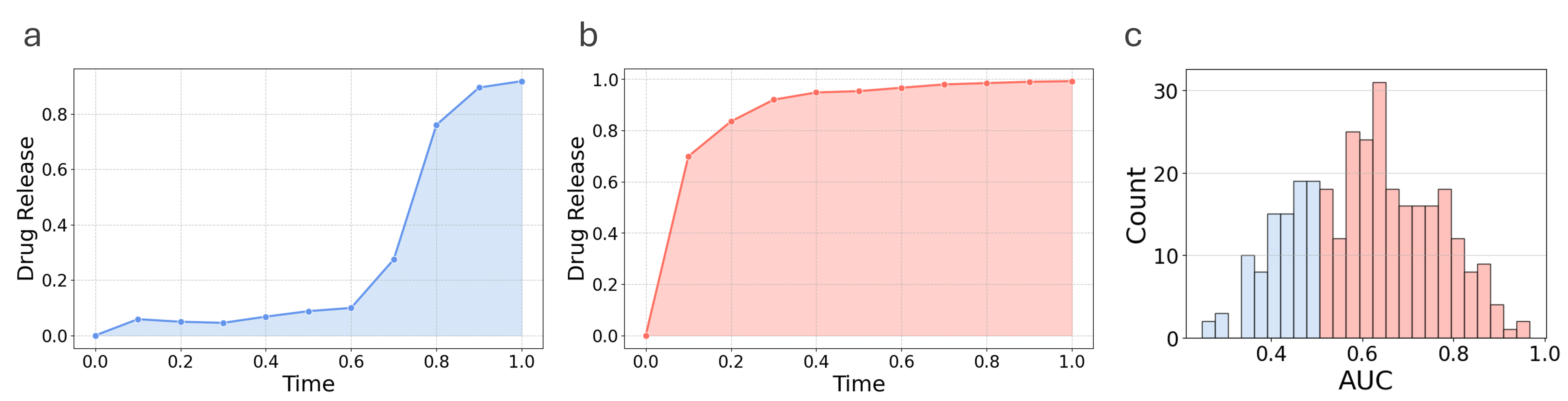}
    \caption{Two distinct types of drug release profiles with a) AUC $\leq$ 0.5 and (b) AUC $>$ 0.5. (c) The sample distribution based on the area under the curve (AUC) of release profiles. Blue for AUC $\leq$ 0.5 and red for AUC $>$ 0.5 (red) profile types.}
    \label{fig:auc_distribution}
\end{figure}

\subsection{Fractional drug release prediction}

We perform a multifaceted experiment to investigate the contribution of LAI material characteristics to drug release.  First, we interpolate fractional drug release at 24, 48, and 72 hours to analyze early drug release. We choose up to 72 hours because it is the minimum duration of drug release profiles in our data set. The first facet of machine learning for drug release is to predict fractional drug release at 24-hour, 48-hour, and 72-hour time points. Regression models in machine learning, including linear regression, random forest (RF), and extreme gradient boosting (XGB). Regression models will reveal the influence of individual material characteristics in the early stage of drug release.


\subsection{Drug release classification}

In the second experiment, we group drug release profiles in the data set on the basis of area under the curve (AUC). An AUC of 0.5 represents a monophasic or linear relationship between the drug release  profile and time. Monophasic, delayed biphasic and triphasic drug release profiles, as defined in~\citep{yoo2020burstrelease}, remain below the toxicity thresholds \emph{in vivo}  and can be represented with AUC $\leq$ 0.5, as shown in Figure\ref{fig:auc_distribution} (a). On the other hand, burst biphasic profiles risk exceeding the toxicity thresholds \emph{in vivo} \citep{yoo2020burstrelease}. A burst biphasic \citep{yoo2020burstrelease} drug release profile is categorized by initial rapid release followed by slower release rates and can be represented by an AUC $>0.5$, as shown in Figure~\ref{fig:auc_distribution}(b). We conduct ML-based classification of two distinct drug release profile types: AUC $\leq$ 0.5 versus AUC $>$ 0.5, and investigate the contributions of material characteristics in differentiating these two profile types. Classifier versions of regression models, such as logistic regression, RF, and XGB, can be considered for this task. It is important to note that data structured in rows (formulations) and columns (material characteristics), as tabular data, are better predicted using traditional machine learning (such as decision tree-based models) than deep learning methods~\citep{Grinsztajn2022, Borisov2022_survey, Abrar2022Perturb}.

\subsection{Prediction of complete drug release profile}

\begin{figure*}[t]
    \centering
    \includegraphics[width=\linewidth]{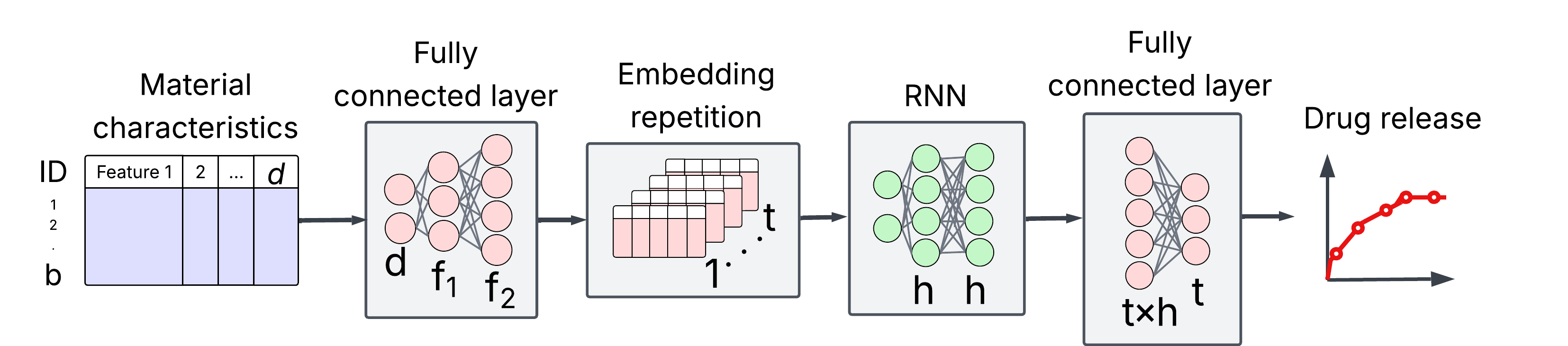}
    \caption{Customized recurrent neural network framework used for complete drug release profile prediction.}
    \label{fig:fc_nn_rnn}
\end{figure*}
Third, we predict the entire drug release profile using the static material characteristics presented in Table \ref{features}. No preceding or partial release data will be used to predict the drug release profile. Time and time-dependent variables are excluded to investigate the individual and collective contributions of material characteristics to a standardized representation of drug release dynamics.
The third facet of machine learning of drug release dynamics learns the relationship between static material characteristics and drug release profile. Recurrent neural networks (RNNs) are specifically designed to learn sequenced data similar to a uniformly sampled release profile. However, basic RNNs suffer from vanishing or exploding gradients, which can cause long-term information to be lost. Long-short-term memory (LSTM)~\citep{hochreiter1997lstm} and gated recurrent units (GRUs)~\citep{cho-gru-2014} address the limitations of basic RNNs. The conventional application of RNNs assumes time-series input data $X\in \Re^{(N \times T\times d)}$ where $N$, $d$, and $T$ represent the number of samples, the number of features, and total points in a sequence, respectively. The output of RNN is $Y\in \Re^{(N\times T\times h)}$ where $h$ is the size of the hidden layer. However, in this study, static material characteristics are used to predict uniformly sampled drug release profile. Therefore, a custom RNN-based framework is developed to learn the relationship between static characteristics and dynamic profiles, as shown in Figure~\ref{fig:fc_nn_rnn}. Input data $X\Re^{N\times d}$ with $d$ static features of $N$ samples are fed into a fully connected neural network (FC-NN) to obtain a new $f$ dimensional feature space $Z\in \Re^{(N\times f)}$. The f-dimensional space is repeated $T$ times to create an input space $ZZ\in\Re^{(N\times T\times f)}$ suitable for RNNs. The output of an RNN is $Y\in\Re^{(N\times T\times h)}$, which is resized to  $Y\in\Re^{((N \times T)\times h)}$ .  The final layer of FC-NN projects $Y$ onto the release profile $\hat{R}\in\Re^{(N \times T\times 1)}$, which is reshaped to $\hat{R}\in\Re^{(N \times T)}$. The RNN is optimized through backpropagation to minimize the sum of the mean squared error (MSE), as shown in Equation \ref{eq:mse}. 


\begin{equation}
\label{eq:mse}    
\text{MSE} = \frac{1}{N} \frac{1}{T} \sum_{i=1}^{N} \sum_{t=1}^{T} (R_{i,t} - \hat{R}_{i,t})^2
\end{equation}

Here, the MSE is between the actual release magnitude ($R$) and the estimated magnitude $\hat{R}$ and averaged across T uniformly sampled points in a drug release profile.

\subsection {Evaluation and explainability} 

\begin{figure}[t]
    \centering
    \includegraphics[width=0.7\linewidth]{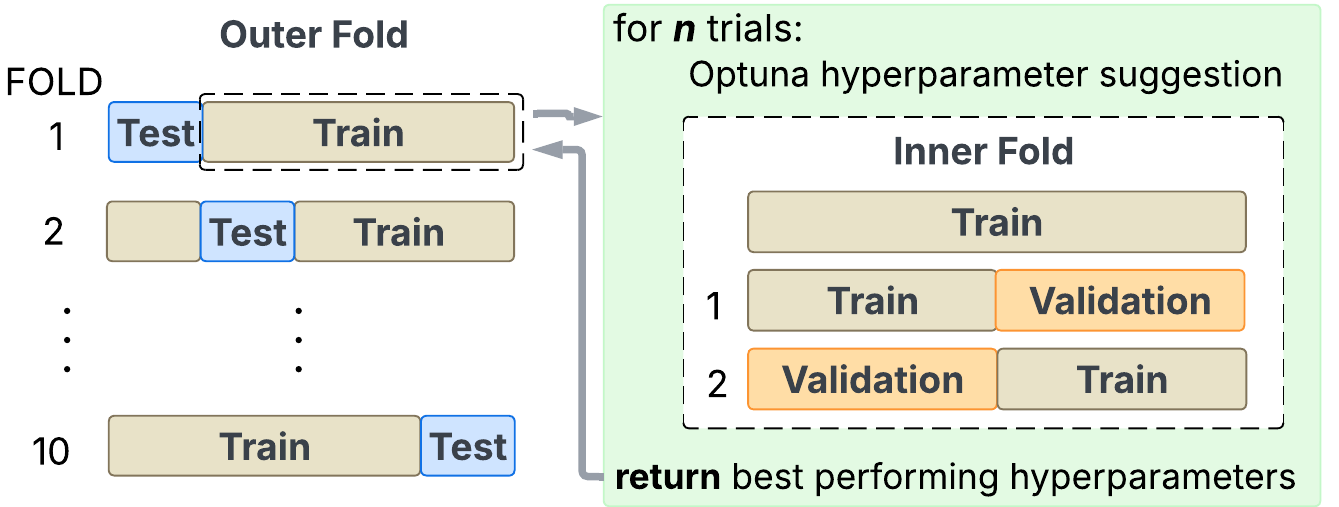}
    \caption{Model validation and selection strategy using 10x2 nested cross-validation.}
    \label{fig:data-split}
\end{figure} 

A 10x2 nested cross-validation scheme is used to train and evaluate all machine learning and RNN models, as shown in Figure~\ref{fig:data-split}. For ten outer folds, one test fold is left out every time. Within the nine training folds, two inner folds are created to tune the model hyperparameters. 
Input features are standardized within the nine folds, which is applied to the test fold. The validation step identifies the best hyperparameter setting for model selection using the average scores across the two inner folds. We run the Optuna package \citep{Akiba2019} for 50 trials to find the best hyperparameter setting. The selected model is trained on nine-fold data and tested on the left-out fold. The average scores across ten left-out folds are reported as model performance. In the fractional drug release prediction task at 24, 48, and 72 hours, the performance of regression models is evaluated using the root mean squared error (RMSE) between the actual ($y$) and predicted ($\hat{y}$) release as below.
\begin{equation}
\label{eq:rmse}    
\text{RMSE} = \sqrt{\frac{1}{n} \sum_{i=1}^{n} (y_i - \hat{y}_i)^2}
\end{equation}
Where \textit{n} represents the number of samples. We also report the Pearson correlation coefficient between the actual and predicted fractional drug release for the three time points. In the classification of drug release profile types, the model performance is reported using average classification accuracy, area under the receiver operating characteristic curve (AUROC), precision, recall, and F1 scores. In the third experiment, the predicted drug release profiles are evaluated using RMSE and correlation scores. The explainability of material characteristics for predicting fractional drug release at 24, 48, and 72 h, and for classifying drug release profile types, is provided using the Shapley additive explanations (SHAP)~\citep{lundberg2017unified} method implemented in the shap Python library. SHAP scores for individual material characteristics quantitatively demonstrate their impact on drug release dynamics.


\section{Results} \label{results}

All experiments are performed on Ubuntu 22.04, powered by an Intel(R) Xeon(R) W-2265 CPU (24 logical cores) running at 3.70 GHz, 64 GB of RAM, and a Quadro RTX A4000 GPU with 16 GB of video memory. The machine learning models are implemented and evaluated using the Sci-Kit learning package in Python. The RNN models are implemented using the PyTorch package in Python, which leverages parallel processing to automatically utilize multiple CPU cores.

\subsection{Data and model preparation}


Machine learning models have a set of hyperparameters that must be appropriately tuned for all prediction tasks. The hyperparameter spaces used for machine learning and RNN models are as follows. The logistic regression model includes the inverse of regularization strength (C: 1e-4 to 1e4), penalty ('l1', 'l2'), solver ('liblinear'), maximum iterations (200 to 1000), class weight ('None', 'balanced'), and tolerance (1e-6 to 1e-3). Random forest hyperparameters include the number of estimators (50 to 300), maximum depth (3 to 20), minimum sample split (2 to 20), minimum samples leaf (1 to 10), maximum features (sqrt, log2), and bootstrap (True, False). The XGB hyperparameters include maximum depth (3 to 20), learning rate (0.01 to 0.3), number of estimators (50 to 300), column subsample by tree (0.5 to 1.0), gamma (0 to 5), regularization alpha (0 to 1), and regularization lambda (0 to 1). For the release profile prediction, RNN hyperparameters include hidden size (32, 64, 128, 256), number of layers (1, 2, 3), dropout (0.2, 0.3, 0.4), learning rate (1e-3, 1e-4), and batch size (16, 32, 64, 128). RNN models are trained for 250 epochs in the inner fold and 500 epochs in the outer fold, based on our observations of the training and validation loss curves.

\subsection{Prediction of fractional drug release}

\begin{table*}[t]
\centering
\caption{Performance scores of regression models in predicting fractional drug release at varying early release time points.}
\label{regressor_table}
\scalebox{0.75}{
\begin{tabular}{lcccccc}
\toprule
&\multicolumn{2}{c}{24 h}&\multicolumn{2}{c}{48 h}&\multicolumn{2}{c}{72 h}\\
\midrule 
\textbf{Model} & \textbf{RMSE} & \textbf{Correlation} & \textbf{RMSE}  & \textbf{Correlation}& \textbf{RMSE} & \textbf{Correlation} \\
\midrule
LinR & 0.18 (0.02) & 0.39 & 0.21 (0.02)  & 0.39 & 0.22 (0.02)   & 0.42 \\
RF & 0.15 (0.03) & 0.65 & 0.17 (0.03) & 0.65 & 0.18 (0.03) & 0.66 \\
XGB & 0.15 (0.03) & 0.63 & 0.17 (0.03) & 0.65 & 0.18 (0.03) & 0.67 \\ 
\bottomrule
\end{tabular}}
\end{table*}

\begin{figure*}[t]
    \centering
    \includegraphics[width=\linewidth]{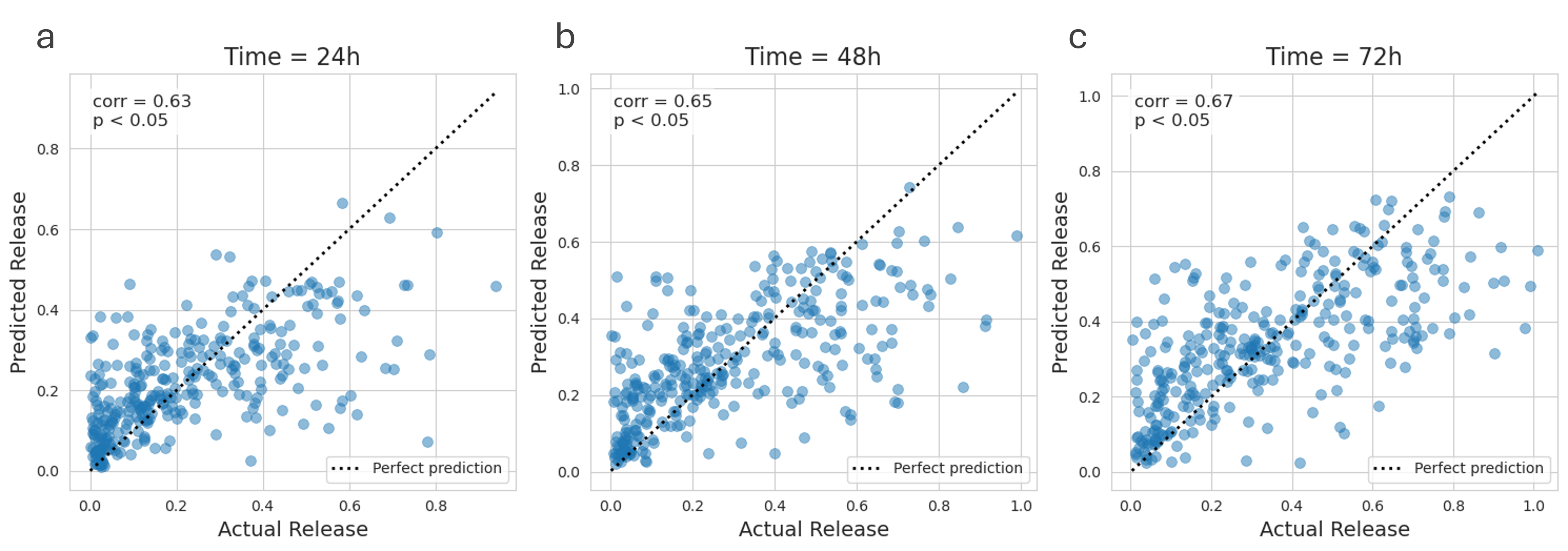}
    \caption{Prediction of the drug release magnitude after (a) 24 hours, (b) 48 hours, and (c) 72 hours following the onset of the drug release using the best regression model (XGB).}
    \label{fig:regressor_results}
\end{figure*}

Fractional drug release values at 24, 48, and 72 hours are predicted from the formulation and microparticle characteristics using regression models to investigate predictors of early drug releas. Table~\ref{regressor_table} presents the regression model performance. Similar to AUC-type classification, non-linear regression models are far superior to their linear counterparts. RMSE scores are relatively higher (better) at an earlier time point (24 h) than at a later time point (72 h). The correlation between true and predicted drug release magnitude is the highest (0.67) at 72 hour point. 

\subsection{Prediction of drug release types}

\begin{table*}[t]
\centering
\caption{Average performance scores with standard deviations in classifying drug release profile types. }
\label{tab:classifier_results}
\begin{tabular}{llllll}
\toprule
\textbf{Model} & \textbf{Accuracy} & \textbf{AU-ROC} & \textbf{Precision} & \textbf{Recall} & \textbf{F1-score} \\  
\midrule 
LR & 0.74 (0.04) & 0.69 (0.09) & 0.75 (0.03) & 0.96 (0.05) & 0.84 (0.03) \\
RF & 0.80 (0.04) & 0.80 (0.08) & 0.83 (0.04) & 0.92 (0.05) & 0.87 (0.03) \\ 
XGB & 0.80 (0.04) & 0.78 (0.07) & 0.84 (0.04) & 0.91 (0.07) & 0.87 (0.03) \\ 
\bottomrule
\end{tabular}
\end{table*}

Table~\ref{tab:classifier_results} presents the drug release type prediction performance of machine learning models. The data set consisted of 321 PLGA-based drug carrier formulations and the corresponding drug release profiles published by Bao et al. \citep{Bao2025}. Figure~\ref{fig:auc_distribution} (c) shows that the distributions of drug release types based on the AUC cutoff are not balanced (AUC $\leq$ 0.5: 84, AUC $>$ 0.5: 237). 

The nonlinear and nonparametric decision tree-based models (RF and XGB) perform significantly better than the linear logistic regression classifier. The difference in prediction performance suggests a nonlinear relationship between the material characteristics and the drug release profiles, which a linear parametric model cannot capture robustly. In general, an 80\% classification accuracy and an F-1 score of 0.87 demonstrate that material characteristics are strongly predictive of drug release profile types.

\begin{table*}
\centering
\caption{Average RMSE between the true and the predicted complete drug release profiles for different release profile types based on AUC.}
\label{RNN_results}
\scalebox{0.85}{
\begin{tabular}{lccc}
\toprule
\textbf{Model} & \textbf{AUC $\leq$ 0.5} & \textbf{AUC $>$ 0.5} & \textbf{Overall}  \\
\midrule
XGB -Time               & 0.140 (0.090) & 0.121 (0.064) & 0.147 (0.010) \\
XGB - No Time           & 0.320 (0.065) & 0.301 (0.055) & 0.314 (0.008) \\
FC-NN-LSTM - No Time    & 0.131 (0.085) & 0.130 (0.069) & 0.148 (0.017) \\
FC-NN-GRU - No Time     & 0.137 (0.094) & 0.130 (0.070) & 0.152 (0.020) \\
\bottomrule
\end{tabular}}
\end{table*}

\subsection{Complete drug release profile prediction}
\begin{figure*}[t]
    \centering
    \includegraphics[width=\linewidth]{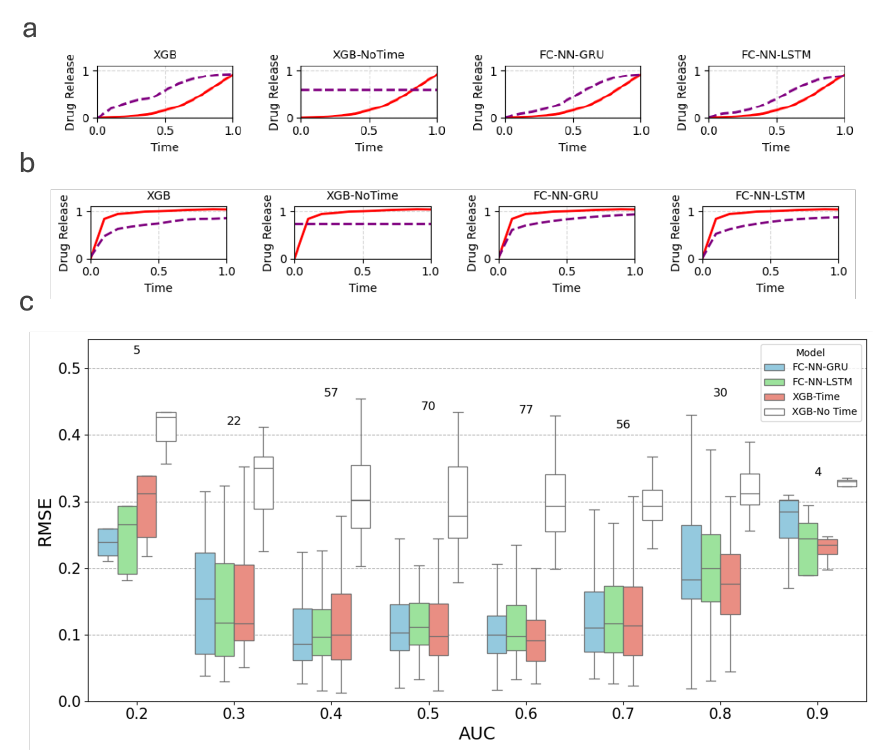}
    \caption{Predicted drug release profiles with (a) AUC $\leq$ 0.5 and (b) AUC $>$ 0.5. (c) Box-and-whisker plot comparing RMSE prediction scores of different models for varying AUC of drug release profiles.}
    \label{fig:profile_prediction}
\end{figure*}

One of the contributions of our study is the development of a custom RNN-based learning framework (Fig. \ref{fig:fc_nn_rnn}) to predict complete drug release profiles from the static characteristics of materials. Previous methods~\citep{Bannigan2023} have reported time as the primary predictor of drug release, which is certainly a time-dependent process. However, time is not an informative predictor because the drug release profile monotonically increases with time. Therefore, our custom RNN framework predicts the complete release profile solely from material characteristics, without using time as an input variable. Table~\ref{RNN_results} presents the performance scores in the prediction of the complete drug release profile. The XGB model, which takes material characteristics and time as inputs, yields the lowest error for release profile with AUC $>$ 0.5. However, both FC-NN-RNNs achieve better performance in predicting release profiles with AUC $\leq$ 0.5. It is important to note that the XGB model fails without the time input, resulting in a flat drug-release profile and, ultimately, a high prediction error. In contrast, RNNs are specifically designed for sequence data. Overall, XGB with time performs marginally better than the proposed RNN, FC-NN-LSTM without time. However, our custom RNN framework with an LSTM, using material characteristics, achieves comparable accuracy, a remarkable result given the absence of the primary time predictor typically used in drug release prediction models. 

Several representative release profiles for the two types are shown in Figure~\ref{fig:profile_prediction}. Figure~\ref{fig:profile_prediction}(a) shows that RNN-based models better capture the low AUC profiles (AUC $\leq$ 0.5) where XGB-Time tends to overestimate. In contrast, RNN models tend to underestimate release profiles with high AUC scores (AUC $>$ 0.5) compared to XGB-Time, as shown in Figure~\ref{fig:profile_prediction}(b). The statistical dependence between the model prediction error and the AUC scores of individual prediction models is presented in a box-and-whisker plot in Figure \ref{fig:profile_prediction}(c). The figure shows that profiles with low AUC scores (AUC $<$ 0.5) are better predicted by RNN-based models. Profiles with AUC = 0.5, representing monophasic of linear release, are predicted comparably by XGB-Time and the FC-RNN models. However, for high AUC scores, the prediction of the drug release profile becomes more time-dependent when the XGB-Time model performs the best. As shown in the same figure, without time information, the XGB model fails to achieve comparable predictive performance. In general, our results show that RNN-based models predict profiles with AUC$\leq$0.5 robustly, suggesting that controlled release can be predicted from static material properties. Furthermore, the drug release curves associated with AUC $\leq$0.5 such as delayed biphasic and triphasic curves, represent more complex drug release mechanisms than diffusion-based burst biphasic curves. FC-NN-RNNs capture these curves robustly from material characteristics alone. Prior drug release prediction studies have not explicitly evaluated model performance on such complex drug release profiles. The predominance of diffusion-driven burst biphasic curves with simpler more predictable kinetics may inflate the accuracy of time-dependent models. Evaluating models across diverse release types, as demonstrated here, provides a more rigorous assessment of predictive capability.

\subsection{Explainable drug release dynamics}

\begin{figure*}[t]
    \centering
    \includegraphics[width=1.0\linewidth]{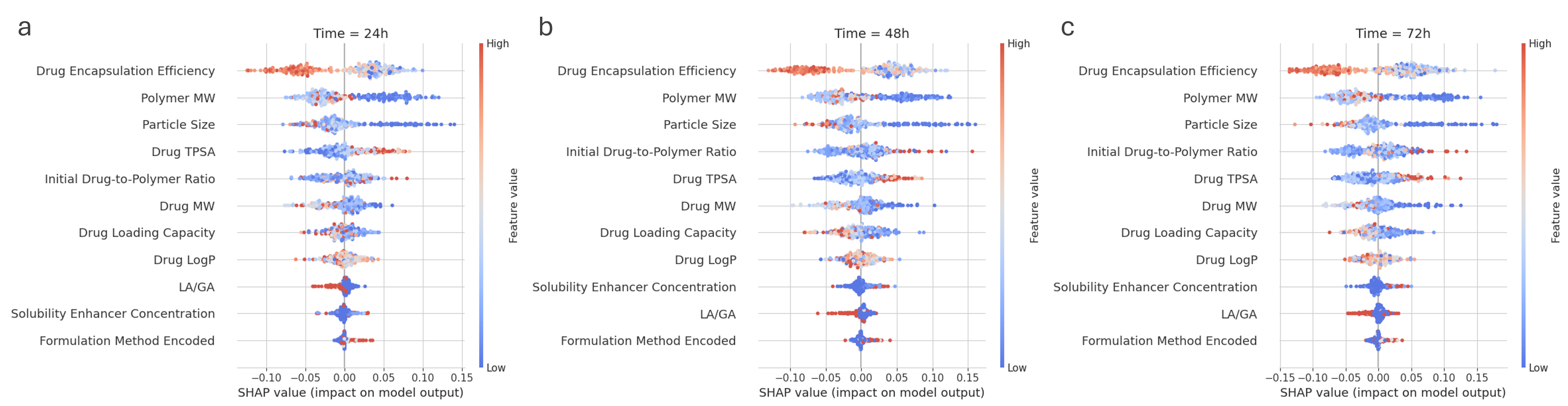}
    \caption{Beeswarm plots for XGB regressor demonstrate SHAP feature impact on model prediction of fractional drug release at (a) 24, (b) 48, and (c) 72 hours of drug release.}
    \label{fig:regressor_results_shap}
\end{figure*}

\begin{figure*}[t]
    \centering
    \includegraphics[width=0.7\linewidth]{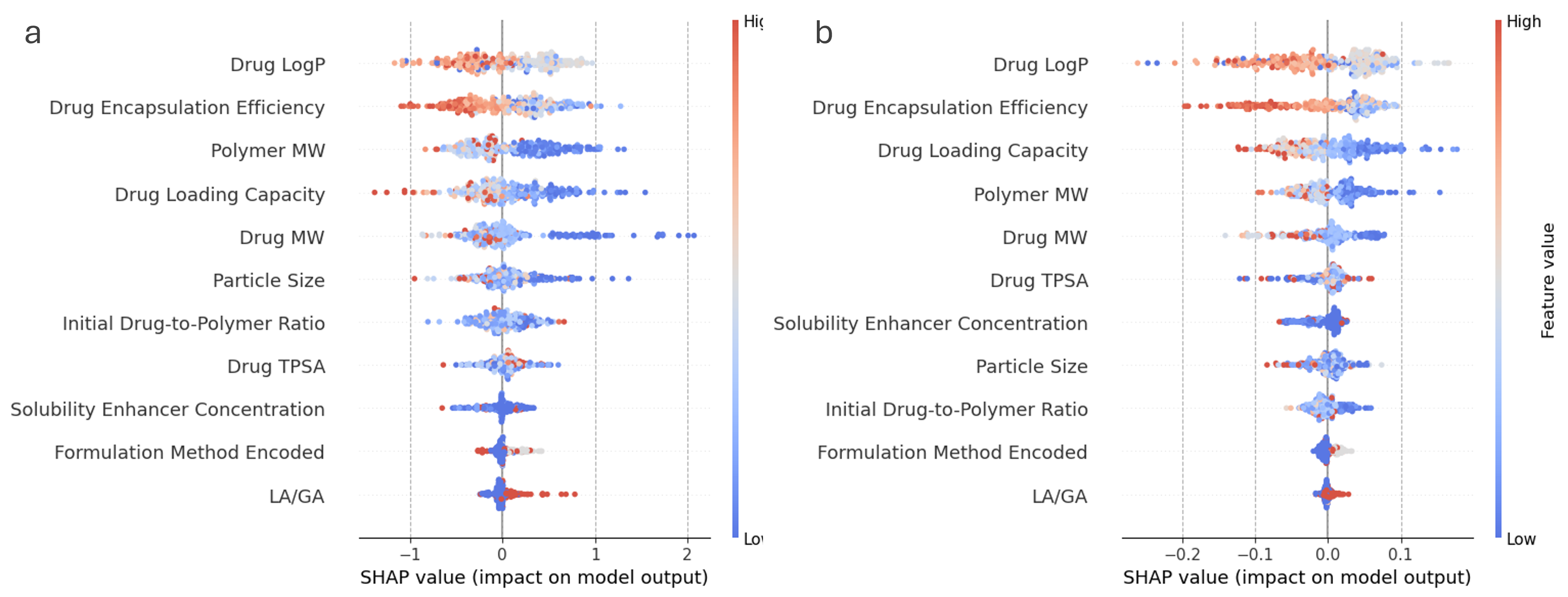}
    \caption{Beeswarm plot for (a) XGB and (b) RF classifiers demonstrate SHAP impact on model prediction of drug release profile type (AUC $\leq$ 0.5: 0, AUC $>$ 0.5 release: 1).} 
    \label{fig:classifier_results_shap}
\end{figure*}

Contributions from material characteristics can further explain the robust prediction of drug-release dynamics. It can be argued that the time dependence of existing predictive models can overshadow the relative importance of material characteristics in drug release dynamics. Our proposed predictive models have achieved promising performance without time as input, suggesting their ability to learn drug-release dynamics from material characteristics.

The effects of individual material characteristics on fractional drug release are demonstrated by the SHAP beeswarm plots in Figure \ref{fig:regressor_results_shap}. For all three time points (24-hour, 48-hour, 72-hour), drug encapsulation efficiency appears as one of the most important characteristics during initial release. Higher drug encapsulation efficiency is associated with lower drug release. Next in the order, lower values of polymer MW and particle size are associated with higher fractional drug release. Higher values of drug TPSA and initial drug-to-polymer ratio are also indicative of higher drug release output. In terms of effect on drug release, drug MW, drug loading capacity, and drug lipophilicity rank next, with lower values of drug MW and loading capacity supporting lower fractional release. However, the effect of drug lipophilicity appears neutral. In general, LA/GA, solubility enhancer concentration, and formulation method demonstrate the lowest influence on drug release.

In classifying drug release profile types, the SHAP beeswarm plots in Figures \ref{fig:classifier_results_shap} a and b show the importance of individual material characteristics and explain their impacts on the prediction of the XGB and RF classifiers, respectively. Both classifiers demonstrate similar rank order of material characteristics in predicting the release profile types. Among the top ranked characteristics, higher values of drug LogP, encapsulation efficiency, and loading capacity are associated with the AUC $\leq 0.5$ release profile type, as reflected in lower output with AUC $\leq 0.5$ (release type = 0) than in release with AUC $>$ 0.5 (release type = 1). In contrast, low Drug MW and Polymer MW values contribute to an increased likelihood of release with AUC $ > 0.5$. The effect of particle size in the same direction is more prominent with the XGB classifier than with the RF classifier. A lower concentration of the solubility enhancer results in lower SHAP values, increasing the likelihood of AUC $<$ 0.5 type release predicted by the RF classifier. In contrast, the effects of formulation methods, drug TPSA, and initial drug-polymer ratio are not prominent in terms of SHAP values.

\section{Discussion} \label{discussion}

The development of LAI involves investigating an optimal combination of materials that yields the desired drug release dynamics, which is an iterative and costly process. To this end, this paper presents one of the first systematic investigations of the complex relationship between material characteristics and drug release dynamics, using data from multiple sources and drug formulations. The findings of the paper can be summarized as follows. First, the material characteristics of polymer-based microparticles contribute substantially to the prediction of early drug release, release profile types, and complete profiles. 
Previous studies have demonstrated this contribution on a limited scale. Second, despite limited samples, the correlation between true and predicted fractional drug release is between 0.63-0.67 within the early phase of drug release (24-72 hours). Third, our methodological contributions demonstrate that material characteristics are predictive of drug release profile types, achieving up to 80\% classification accuracy. Fourth, one of the first efforts to predict an entire drug release profile demonstrates that an RNN-based framework is a better alternative to the time-dependent XGB model used in the literature in terms of accuracy and material contributions, especially in predicting more complex release profiles with AUC$<$0.5, such as delayed biphasic and triphasic curves. 
Fifth, explainable ML methods reveal the strong and consistent effects of individual material characteristics on predicting early fractional drug release values and profile types.

\subsection {Practical descriptors of drug release dynamics}

 The prediction of drug-release profiles of different types with varying release durations and inconsistent time points is a unique problem for data science. Despite widespread applications, traditional ML methods are not optimally designed to learn dynamic profiles from static variables. Current methods for predicting drug release profiles use an ad hoc time-based feature alongside material characteristics \citep{Zawbaa2016, Salma2021, Husseini2025, Deng2023, Bannigan2023, Zhang2025} and demonstrate promising predictive performance. However, time-dependent predictive models conceal the importance of material characteristics that underpin drug release dynamics known from \emph{in-vitro} studies. Therefore, defining suitable target descriptors of drug release dynamics is a critical requirement for effective ML and a key contribution of this paper. Our time-independent data transformation and categorization of drug release profiles have enabled ML across heterogeneous formulations with promising prediction performance and discovery of strong influence of material characteristics. In addition to predicting a global descriptor (AUC) of drug release profiles, we demonstrate the impact of material characteristics on early time points of drug release, which may facilitate LAI optimization to prevent burst release.

\subsection{Effects of LAI material characteristics}

Drug release is controlled by complex intermolecular interactions between the drug, the carrier, and the environment. In the literature, intermolecular interactions between the drug and the carrier have been reported to affect the distribution of the drug within the carrier \citep{park2019injectable,baumann2024prolonged, yoo2020burstrelease}, drug loading \citep{Li03042015}, and the rate of drug release \citep{park2019injectable}. As a result of intermolecular interactions, hydrophobic drugs diffuse through the polymer matrix, and hydrophilic drugs are released primarily through diffusion through pores \citep{yoo2020burstrelease}. Therefore, material characteristics drive intermolecular interactions, which, in turn, determine drug release dynamics. However, the material characteristics of LAIs that drive intermolecular interactions are highly interdependent, which makes prediction of drug release and understanding of the relative contribution of individual characteristics \emph{in-vitro} lab settings challenging. 

Our proposed data transformation and explainable ML approaches address the limitations of previous studies to demonstrate the collective and relative effects of the individual material characteristics of polymer-based LAI on drug release. Most importantly, our method reveals a substantial predictive influence of material characteristics without being overshadowed by time or drug release after a certain time. Table \ref{tab:discussion_summary} explains the rank and relative importance of material characteristics in different studies and contexts, which can help formulation scientists accelerate rational design and save valuable development time by focusing on the most important ones. Our results using explainable ML identify the leading predictors of early drug release between 24-72 hours and predictors that differentiate between drug release profiles with low versus high AUC. We provide further insights into the effects of these predictors in the following sections.

\subsubsection{Effects on early drug release}
Table \ref{tab:discussion_summary} shows drug encapsulation efficiency, polymer MW, particle size, drug TPSA, and initial drug-to-polymer ratio are the leading predictors of early-stage drug release across all three time points. Higher values of drug encapsulation efficiency, polymer MW, and particle size are associated with lower early-stage release. 

\textbf{Encapsulation efficiency.} Encapsulation efficiency refers to the percentage of initial drug encapsulated in the carrier which can be attributed to carrier fabrication influenced by intermolecular interactions~\citep{Li03042015, Bickerton2012}. In a recent ML study \citep{Bannigan2023}, the encapsulation efficiency was not reported for the prediction of drug release. However, our findings show that high values of drug encapsulation efficiency predict lower early release. Encapsulation efficiency can help describe drug-carrier stability, as high values of encapsulation efficiency indicate successful encapsulation and support low early-stage release \citep{LUAN2006168}, as supported by our findings. 

\textbf{Polymer MW.} Our analysis also identifies high polymer molecular weight as a top predictor of low early-stage release. It has been reported in \citep{Feng2015EffectsOD} that high-MW polymers have longer polymer chains, leading to chain entanglements and reduced water influx, thus reducing polymer degradation and drug diffusion rates \citep{Feng2015EffectsOD}. Our results confirm that high values of polymer MW predict low early-stage release. Moreover, our results rank polymer MW as one of the leading predictors when considering numerous formulation variables.

\textbf{Particle size.} Furthermore, our findings identify smaller particle size as associated with increased early release. In the literature, one study reported that larger particle size was associated with higher drug release rates \citep{siepmann2004effect}, while another reported results contrary to this observation \citep{klose2006porosity}. Our results show larger particles result in slower release in the early stages of release. 

\textbf{Drug TPSA.} Drug descriptors such as topological polar surface area (TPSA) are intrinsically related to other molecular descriptors such as drug lipophilicity and molecular weight, and their influence is difficult to isolate or manipulate in \emph{in-vitro} studies as a result. For example, drug analogs with different TPSA but identical MW or LogP values do not exist. Our results demonstrate high drug polarity (TPSA) is associated with increased early drug release prediction. Drugs with high TPSA are highly soluble in water and released primarily through diffusion, while prolonged release of non-polar drugs is supported through drug-polymer affinity and erosion-based release mechanisms. Importantly, these inferences are dependent on high encapsulation efficiency, which was a primary predictor across all models.

\textbf{Initial drug-to-polymer ratio.} The influence of initial drug-to-polymer ratio is also ranked among the top 5 for early drug release with the highest values being associated with increased fractional release. The initial drug-to-polymer ratio influences drug distribution throughout the polymer matrix. It is a formulation variable related to encapsulation efficiency and drug loading capacity. For example, with high drug-to-polymer ratio, particles can become overloaded and result in poor encapsulation and uncontrolled release.

\subsubsection{Effects on drug release profile types}

\begin{table*}
\centering
\caption{Comparison of factors important for predicting early drug release and drug release profile types using polymer-based LAIs. Delayed release indicates low early release and drug release profiles with AUC $<0.5$.
N.R.: not reported; N.S.: not significant. N.E.: neutral effect.}
\label{tab:discussion_summary}
\scalebox{0.7}{
\begin{tabular}{>{\raggedright\arraybackslash}m{0.25\linewidth} >{\raggedright\arraybackslash}m{0.25\linewidth} >{\raggedright\arraybackslash}m{0.25\linewidth} >
{\raggedright\arraybackslash}m{0.25\linewidth} >
{\raggedright\arraybackslash}m{0.25\linewidth}}
\toprule
\textbf{Factors} & \textbf{\textit{in-vitro}} & \textbf{Baseline } & \textbf{Early drug} & \textbf{Drug release}  \\
&&{\bf ML study} \citep{Bannigan2023}&{\bf release}& {\bf profile types}\\
\midrule
Drug LogP & $\uparrow$ Delayed release \citep{baumann2024prolonged}  & N.S. & N.E. & $\uparrow$ Delayed release \\
\addlinespace
Encapsulation efficiency & $\uparrow$ Delayed release \citep{LUAN2006168} & N.R. & $\uparrow$ Delayed release & $\uparrow$ Delayed release \\
\addlinespace
Polymer MW & $\uparrow$ Delayed release \citep{Feng2015EffectsOD} & $\uparrow$ Delayed release & $\uparrow$ Delayed release & $\uparrow$ Delayed release \\ 
\addlinespace
Drug TPSA & N.R. & $\uparrow$ Delayed release & $\downarrow$ Delayed release & N.E. \\
\addlinespace
Loading capacity & $\uparrow$ Delayed release \citep{Mao2007EffectOW} \newline $\downarrow$ Delayed release \citep{Kim1992EffectOL, Bickerton2012} & N.S. & $\uparrow$ Delayed release & $\uparrow$ Delayed release \\
\addlinespace
Initial DPR 
& $\downarrow$ Delayed release \citep{Mishra_Waghamare_Khanage_2025} & N.E. & $\downarrow$ Delayed release & N.E. \\
\addlinespace
Drug MW & $\downarrow$ Delayed release \citep{Mao2007EffectOW} \newline $\uparrow$ Delayed release \citep{alhnan2010inhibiting} & $\uparrow$ Delayed release & $\uparrow$ Delayed release & $\uparrow$ Delayed release \\ 
\addlinespace
Particle size & $\downarrow$ Delayed release \citep{siepmann2004effect}  \newline $\uparrow$ Delayed release \citep{klose2006porosity} & N.R. & $\uparrow$ Delayed release & $\uparrow$ Delayed release \\
\midrule 
\addlinespace
Top 5 (rank) & Isolated effects. & Time, release at\newline 24 h, Drug MW,\newline Polymer MW, Drug TPSA & Encapsulation efficiency, Polymer MW, Particle size, Drug TPSA, Initial DPR 
& Drug LogP, \newline  Encapsulation efficiency, Polymer MW, Loading capacity, Drug MW  \\
\midrule 
\addlinespace
Relative importance &  Relative contribution \newline  is unknown   & Time-related \newline variables account \newline  for  $>$50\% of \newline SHAP values.  & Top 3 material 
\newline characteristics \newline account for  30\% of SHAP values &  Top 3 material 
\newline characteristics \newline account for  30\% of SHAP values\\  
\bottomrule 
\end{tabular}}
\end{table*}


Table \ref{tab:discussion_summary} shows that drug lipophilicity, loading capacity, encapsulation efficiency, and drug and polymer molecular weight are the primary predictive material characteristics of drug release profile types. Our findings suggest that there are some similarities and differences between predictors of early drug release and drug release profile types. In both early release and profile types, high values of encapsulation efficiency and polymer MW demonstrate strong influence in delaying the release. 

\textbf{Drug lipophilicity.} Drug lipophilicity  (LogP) refers to the ability of the drug molecule to dissolve in lipid versus aqueous environments. Drug lipophilicity also impacts intermolecular drug-carrier and drug-environment interactions to the extent that increasing lipophilicity can change controlled release from burst to delayed release, and lipophilic or fat-soluble drugs tend to release more slowly than water-soluble drugs~\citep{baumann2024prolonged}. However, the relative importance of drug lipophilicity among other material characteristics has remained unclear in the literature. For the first time in an ML study, our results demonstrate that high drug lipophilicity is a primary predictor of delayed drug release. We attribute the effects of increasing drug lipophilicity on delayed release to a transition in release mechanism by which the drug is released. For example, drugs with low lipophilicity values are released more easily by diffusion and thus associated with burst biphasic curves in the AUC $>$ 0.5 class. On the other hand, drugs with high lipophilicity remain entraped until the polymer erodes represented in delayed biphasic or triphasic release profiles with an AUC $\leq$ 0.5.

\textbf{Drug content descriptors.} Our results identify high values of descriptors of drug-content as some of the top predictors of delayed release, namely drug encapsulation efficiency and loading capacity, both of which can be attributed to carrier fabrication influenced by intermolecular interactions~\citep{Li03042015, Bickerton2012}. Our findings show that high values of drug encapsulation efficiency and loading capacity significantly predict delayed release. It should be noted that, although increasing the loading capacity can aid in delayed release \citep{Mao2007EffectOW},  carriers can become overloaded \citep{Bickerton2012}. Therefore, high loading capacity alone does not necessarily mitigate burst release \citep{Kim1992EffectOL, Bickerton2012}. Together, both drug content descriptors describe the amount of drug available and the stability of the encapsulation and are important predictors of delayed release.

\textbf{Molecular weight.} Our analysis also identifies high polymer and drug molecular weights as the top predictors of delayed drug release. Molecular weight (MW) refers to the weight of drug and polymer molecules, which is known to influence intermolecular interactions that regulate drug release. Our results confirm high values of polymer MW predict delayed release in early stages and in the overall drug release profile. Lower drug MW values can represent smaller drugs that are released through pores where larger drug molecules remain entrapped and take longer to diffuse. However, \textit{in-vitro} studies have reported conflicting findings on the effect of drug MW on delayed release \citep{alhnan2010inhibiting, Mao2007EffectOW}. This may be attributed to the limited scope of each study. Our results across diverse drug formulations suggest that high drug MW predicts delayed release. In an explainable ML study~\citep{Bannigan2023}, polymer and drug MW are identified as the main predictors after time-related variables to predict the drug release profile. However, the magnitude of importance of these characteristics remains minimum because of the overpowering influence of time and early drug release information in their ML model.

\textbf{Secondary predictors. }The other material characteristics in our study that merit discussion include particle size, drug TPSA, lactide-to-glycolide ratio, (LA\textbackslash GA) and formulation method. The influence of smaller particle size on uncontrolled drug release ranks higher for early drug release than for the overall profile. This can be attributed to a higher surface-area-to-volume ratio and a shorter diffusion path length, both of which facilitate initial drug release.  While particle size remains a predictor of overall release, its influence is attenuated in cumulative release profiles. We speculate that the change in the behavior of the material characteristics is due to other competing factors that more decidedly predict drug release, according to Shapley additive explanations. In contrast to previous work on ML \citep{Bannigan2023}, our results do not find a significant predictive value for the topological surface area (TPSA) of a drug for the prediction of drug release type, despite its importance for early release prediction. The predictive importance of TPSA for early release, but not overall profile type, likely reflects a transition in the release mechanism from diffusion-based to erosion-controlled. Furthermore, a high lactide-to-glycolide ratio (LA\textbackslash GA)  is recommended to delay the release to reduce the likelihood of burst release\citep{yoo2020burstrelease}. 
In contrast, the LA/GA ratio had a limited influence in our study compared to other factors on the prediction of delayed drug release. The limited influence of LA/GA can also be attributed to the considerable diversity of material characteristics in our study, particularly the wide spectrum of drug lipophilicity. Consequently, the magnitude of influence of the LA/GA ratio may be overshadowed by the predictive strength of other factors. Finally, our results suggest that the oil-in-water emulsion technique, typically used to encapsulate hydrophobic drugs, tends to predict delayed release.

\subsection {Recommendations for materials scientists}

The challenges in optimizing polymer-based long-acting injectables (LAI) arise from the complex and abundant underlying physicochemical processes driving microparticle formation \citep{park2019injectable} and drug release \citep{fredenberg2011mechanisms}. Studies on LAI formulations are growing and many variables have been shown to control drug release. However, due to constraints in laboratory experiments, each study is generally limited in scope to one or two material characteristics \citep{siepmann2004effect, klose2006porosity, baumann2024prolonged}. As a result, the combined contributions of material characteristics to drug release are not fully understood. Even when several material characteristics are considered for drug release experimentation \citep{Mao2007EffectOW}, individual contribution of characteristics and complex relationship in between remain unknown. Therefore, researchers have recently explored ML in the prediction of drug release \citep{Zawbaa2016, Salma2021, Husseini2025, Deng2023, Bannigan2023, Zhang2025} to investigate the contribution of individual characteristics to optimize drug release. This study advances knowledge in the field by addressing the limitations of previous studies to offer some recommendations on how to better control drug release with long-acting formulations.

We offer the following recommendations, which may serve as initial screening criteria rather than strict cutoffs. Researchers developing delayed release formulations should begin with materials selection, opting for hydrophobic candidates with higher (positive) values of drug lipophilicity (LogP: $\sim 1$\text{-}$5$ or higher) or low TPSA ($<$ 60–90 Å²) and medium to high drug ($>$ 300-400 Da or higher) and polymer ($\sim$50-100 kDa or higher) MW.  LAI formulations should be optimized to maximize loading capacity ($\sim$10-20\%) and encapsulation efficiency (approx. $>$ 70\%) by screening the initial drug-to-polymer ratio prior to the characterization of the drug release profile.  Notably, these general recommendations are not exclusive or exhaustive. For example, burst release can be reduced for a hydrophilic or low MW drug by increasing the encapsulation efficiency, polymer MW, or particle size.

Our findings recommend minimizing burst release by substituting or modifying drug candidates with higher lipophilicity and by selecting drug and polymer candidates with medium to high molecular weight. Secondly, optimizing both loading capacity and encapsulation efficiency by tuning the initial drug-to-polymer ratio is critical for controlling drug release. In addition, strategies for further fine-tuning the drug release profile should begin with encapsulation efficiency, drug and polymer molecular weight, followed by particle size.

\subsection{Limitations}

Our study is bounded by data and methodological constraints that should be considered. First, the early time points 24, 48, and 72 hours are chosen on the basis of information about burst release scattered in the literature. Unfortunately, there is no standard criterion for burst release due to the heterogeneity of drug types and therapeutic requirements. Therefore, our data-driven results do not imply the prediction of burst release. Instead, our results should be interpreted as a general recommendation to prevent burst release and delay the drug release. Although time-independent representation of drug release profiles facilitates ML and explainability of material characteristics, time information is still relevant for the optimization of precision drug delivery in time. Moreover, our method uses a data set with limited samples from heterogeneous sources, which may not model drug-specific requirements.

\section{Conclusion} 
\label{conclusion}

This paper establishes a time-independent and explainable machine learning framework that  identifies primary influencers of drug release from long-acting injectables. Our analysis reveals five key predictors of early drug release: encapsulation efficiency, polymer MW, particle size, drug TPSA, and initial drug-to-polymer ratio, and of drug release type: drug lipophilicity (LogP), loading capacity, encapsulation efficiency, and drug and polymer molecular weight. We demonstrate agreement with isolated effects reported in in-vitro studies and present the relative importance of material characteristics driving early release and drug release types for the first time. Importantly, we demonstrate that more complex delayed biphasic and triphasic curves are predicted robustly with material characteristics alone, even though they are less represented in the available data. This finding suggests that the performance of time-dependent models may be inflated by the data presently available, which consists largely of  diffusion-based burst biphasic curves. For the first time, the complete prediction of drug release from polymer-based microparticles is achieved without a time-input feature. This work provides a practical tool for LAI development in which drug release profiles can be predicted according to the characteristics of the material.

\section*{Acknowledgments}
This work was supported by the National Institutes of Health, Grant Number: U54CA163066 and TSUFI-PREM NSF DMR-2424463. 
We thank Ibna Kowsar for support during initial code development.

\bibliographystyle{elsarticle-num} 
\bibliography{ref}

@incollection{xing2020industrialAI,
  author    = {Xing, Feng and Peng, G. (Alex) and Zhang, Bo and Zuo, Shuai and Tang, Jiaqi and Li, Song},
  title     = {Driving Innovation with the Application of Industrial AI in the R\&D Domain},
  booktitle = {Lecture Notes in Computer Science (Including Subseries Lecture Notes in Artificial Intelligence and Lecture Notes in Bioinformatics)},
  volume    = {12203},
  year      = {2020},
  publisher = {Springer},
  doi       = {10.1007/978-3-030-50344-4_18}
}

@inproceedings{adetunla2024AIinMaterials,
  author    = {Adetunla, A. and Akinlabi, E. and Jen, T. C. and Ajibade, S.-S.},
  title     = {Harnessing the Power of Artificial Intelligence in Materials Science: An Overview},
  booktitle = {2024 International Conference on Science, Engineering and Business for Driving Sustainable Development Goals (SEB4SDG)},
  year      = {2024},
  pages     = {1--6},
  address   = {Omu-Aran, Nigeria},
  doi       = {10.1109/SEB4SDG60871.2024.10630185}
}

@article{yoo2020burstrelease,
  author    = {J. Yoo and Y. Y. Won},
  title     = {Phenomenology of the Initial Burst Release of Drugs from PLGA Microparticles},
  journal   = {ACS Biomaterials Science \& Engineering},
  year      = {2020},
  doi       = {10.1021/acsbiomaterials.0c01228}
}

@article{bao2025datasetPLGAMPs,
  author    = {Bao, Zeqing and Kim, Jongwhi and Kwok, Candice and Le Devedec, Frantz and Allen, Christine},
  title     = {A dataset on formulation parameters and characteristics of drug-loaded PLGA microparticles},
  journal   = {Scientific Data},
  year      = {2025},
  volume    = {12},
  number    = {1},
  pages     = {364},
  doi       = {10.1038/s41597-025-04621-9},
  publisher = {Nature Publishing Group}
}

@article{sultana2022nano,
  author    = {A. Sultana and M. Zare and V. Thomas and T. S. S. Kumar and S. Ramakrishna},
  title     = {Nano-based drug delivery systems: Conventional drug delivery routes, recent developments and future prospects},
  journal   = {Medical Innovations \& Development},
  year      = {2022},
  doi       = {10.1016/j.medidd.2022.100134}
}

@article{obrien2021challenges,
  author    = {M. N. O’Brien and W. Jiang and Y. Wang and D. M. Loffredo},
  title     = {Challenges and opportunities in the development of complex generic long-acting injectable drug products},
  journal   = {Journal of Controlled Release},
  year      = {2021},
  doi       = {10.1016/j.jconrel.2021.06.017}
}

@article{elTanani2025revolutionizing,
  author    = {Mohamed El-Tanani and Shakta Mani Satyam and Syed Arman Rabbani and Yahia El-Tanani and Alaa A. A. Aljabali and Ibrahim Al Faouri and Abdul Rehman},
  title     = {Revolutionizing Drug Delivery: The Impact of Advanced Materials Science and Technology on Precision Medicine},
  journal   = {Pharmaceutics},
  year      = {2025},
  volume    = {17},
  number    = {3},
  pages     = {375},
  doi       = {10.3390/pharmaceutics17030375}
}

@article{noorain2023mlplga,
  author    = {L. Noorain and V. Nguyen and H. W. Kim and L. T. B. Nguyen},
  title     = {A Machine Learning Approach for PLGA Nanoparticles in Antiviral Drug Delivery},
  journal   = {Pharmaceutics},
  year      = {2023},
  volume    = {15},
  number    = {2},
  pages     = {495},
  doi       = {10.3390/pharmaceutics15020495}
}

@article{Rezvantalab2024,
  author    = {S. Rezvantalab and S. Mihandoost and M. Rezaiee},
  title     = {Machine learning assisted exploration of the influential parameters on the PLGA nanoparticles},
  journal   = {Scientific Reports},
  volume    = {14},
  pages     = {1114},
  year      = {2024},
  doi       = {10.1038/s41598-023-50876-w}
}

@article{deAzevedo2017,
  author    = {Cristiana Rodrigues de Azevedo and Moritz von Stosch and Mariana S. Costa and A. M. Ramos and M. Margarida Cardoso and Fabienne Danhier and Véronique Préat and Rui Oliveira},
  title     = {Modeling of the burst release from PLGA micro- and nanoparticles as function of physicochemical parameters and formulation characteristics},
  journal   = {International Journal of Pharmaceutics},
  year      = {2017},
  doi       = {10.1016/j.ijpharm.2017.08.118}
}

@article{Bannigan2023,
  author    = {Pauric Bannigan and Zeqing Bao and Riley J. Hickman and Matteo Aldeghi and Florian H{\"a}se and Al{\'a}n Aspuru-Guzik and Christine Allen},
  title     = {Machine learning models to accelerate the design of polymeric long-acting injectables},
  journal   = {Nature Communications},
  volume    = {14},
  pages     = {35},
  year      = {2023},
  doi       = {10.1038/s41467-022-35343-w}
}

@article{Aghajanpour2025,
  author    = {Sareh Aghajanpour and Hamid Amiriara and Mehdi Esfandyari-Manesh and Pedram Ebrahimnejad and Haziq Jeelani and Andreas Henschel and Hemant Singh and Rassoul Dinarvand and Shabir Hassan},
  title     = {Utilizing machine learning for predicting drug release from polymeric drug delivery systems},
  journal   = {Computers in Biology and Medicine},
  volume    = {188},
  pages     = {109756},
  year      = {2025},
  issn      = {0010-4825},
  doi       = {10.1016/j.compbiomed.2025.109756}
}

@article{Zawbaa2016,
  author    = {Hossam M. Zawbaa and Jakub Szlek and Cristian Grosan and Rafal Jachowicz and Aleksander Mendyk},
  title     = {Computational Intelligence Modeling of the Macromolecules Release from PLGA Microspheres—Focus on Feature Selection},
  journal   = {PLOS ONE},
  volume    = {11},
  number    = {6},
  pages     = {e0157610},
  year      = {2016},
  doi       = {10.1371/journal.pone.0157610}
}

@article{Deng2023,
  author    = {Jie Deng and Zhijiang Ye and Wenhao Zheng and Jing Chen and Huan Gao and Ziyang Wu and Grace Chan and Yitao Wang and Dan Cao and Yichun Wang and Simon M. Lee and Dexi Ouyang},
  title     = {Machine learning in accelerating microsphere formulation development},
  journal   = {Drug Delivery and Translational Research},
  volume    = {13},
  number    = {4},
  pages     = {966--982},
  year      = {2023},
  month     = apr,
  doi       = {10.1007/s13346-022-01253-z}
}

@article{Salma2021,
  author    = {Hanan Salma and Yacine Melha and Lounis Sonia and Hamza Hamza and Nourredine Salim},
  title     = {Efficient Prediction of In Vitro Piroxicam Release and Diffusion From Topical Films Based on Biopolymers Using Deep Learning Models and Generative Adversarial Networks},
  journal   = {Journal of Pharmaceutical Sciences},
  volume    = {110},
  number    = {6},
  pages     = {2531--2543},
  year      = {2021},
  month     = jun,
  doi       = {10.1016/j.xphs.2021.01.032}
}

@article{Husseini2025,
  author    = {Ghaleb A. Husseini and Rania Sabouni and Vladislav Puzyrev and Mehdi Ghommem},
  title     = {Deep Learning for the Accurate Prediction of Triggered Drug Delivery},
  journal   = {IEEE Transactions on NanoBioscience},
  volume    = {24},
  number    = {1},
  pages     = {102--112},
  year      = {2025},
  month     = jan,
  doi       = {10.1109/TNB.2024.3426291}
}

@article{Meyer2022,
  author    = {Travis A. Meyer and Cesar Ramirez and Matthew J. Tamasi and Adam J. Gormley},
  title     = {A User’s Guide to Machine Learning for Polymeric Biomaterials},
  journal   = {ACS Polymers Au},
  volume    = {3},
  number    = {2},
  pages     = {141--157},
  year      = {2022},
  doi       = {10.1021/acspolymersau.2c00037}
}

@article{alqarni2025predicting,
  author    = {Alqarni, S. and Huwaimel, B.},
  title     = {Predicting PLGA nanoparticle size and zeta potential in synthesis for application of drug delivery via machine learning analysis},
  journal   = {Scientific Reports},
  volume    = {15},
  pages     = {20765},
  year      = {2025},
  doi       = {10.1038/s41598-025-06872-3}
}

@article{Bao2025,
  title = {Polymer microparticles in an evolving drug delivery landscape: challenges and the role of machine learning},
  author = {Bao, Zeqing and Kim, Jongwhi and Le Devedec, Frantz and Clasky, Aaron and Allen, Christine},
  journal = {International Journal of Pharmaceutics},
  volume = {682},
  year = {2025},
  pages = {125906},
  issn = {0378-5173},
  doi = {10.1016/j.ijpharm.2025.125906},
  url = {https://www.sciencedirect.com/science/article/pii/S0378517325007434},
}

@article{Zhang2025,
  author = {Zhang, Zhenyu and Zhang, Biao and Chen, Rui and Zhang, Qian and Wang, Kai},
  title = {The Prediction of the In Vitro Release Curves for PLGA-Based Drug Delivery Systems with Neural Networks},
  journal = {Pharmaceutics},
  year = {2025},
  volume = {17},
  number = {4},
  pages = {513},
  doi = {10.3390/pharmaceutics17040513},
  pmid = {40284508},
  pmcid = {PMC12030581},
  publisher = {MDPI},
  note = {Published April 14, 2025}
}

@article{baumann2024prolonged,
  title={Prolonged release from lipid nanoemulsions by modification of drug lipophilicity},
  author={Baumann, Nina and Baumgarten, Janosch and Kunick, Conrad and Bunjes, Heike},
  journal={Journal of Controlled Release},
  volume={374},
  pages={478--488},
  year={2024},
  publisher={Elsevier}
}

@article{fredenberg2011mechanisms,
  title={The mechanisms of drug release in poly (lactic-co-glycolic acid)-based drug delivery systems—A review},
  author={Fredenberg, Susanne and Wahlgren, Marie and Reslow, Mats and Axelsson, Anders},
  journal={International journal of pharmaceutics},
  volume={415},
  number={1-2},
  pages={34--52},
  year={2011},
  publisher={Elsevier}
}

@article{Grinsztajn2022,
archivePrefix = {arXiv},
arxivId = {2207.08815},
author = {Grinsztajn, L{\'{e}}o and Oyallon, Edouard and Varoquaux, Ga{\"{e}}l},
eprint = {2207.08815},
journal = {arXiv preprint arXiv:2207.08815},
month = {jul},
title = {{Why do tree-based models still outperform deep learning on tabular data?}},
url = {http://arxiv.org/abs/2207.08815},
year = {2022}
}

@article{Borisov2022_survey,
  title={Deep neural networks and tabular data: A survey},
  author={Borisov, Vadim and Leemann, Tobias and Se{\ss}ler, Kathrin and Haug, Johannes and Pawelczyk, Martin and Kasneci, Gjergji},
  journal={IEEE Transactions on Neural Networks and Learning Systems},
  year={2022},
  publisher={IEEE}
}

@article{Abrar2022Perturb,
author = {Abrar, Sakib and Samad, Manar D.},
doi = {10.1016/j.neunet.2022.09.020},
issn = {08936080},
journal = {Neural Networks},
month = {dec},
pages = {160--169},
publisher = {Elsevier BV},
title = {{Perturbation of deep autoencoder weights for model compression and classification of tabular data}},
volume = {156},
year = {2022}
}

@inproceedings{Akiba2019,
author = {Akiba, Takuya and Sano, Shotaro and Yanase, Toshihiko and Ohta, Takeru and Koyama, Masanori},
title = {Optuna: A Next-generation Hyperparameter Optimization Framework},
year = {2019},
isbn = {9781450362016},
publisher = {Association for Computing Machinery},
address = {New York, NY, USA},
doi = {10.1145/3292500.3330701},
booktitle = {Proceedings of the 25th ACM SIGKDD International Conference on Knowledge Discovery and Data Mining},
pages = {2623–2631},
numpages = {9},
}

@inproceedings{cho-gru-2014,
    title = "Learning Phrase Representations using {RNN} Encoder{--}Decoder for Statistical Machine Translation",
    author = {Cho, Kyunghyun  and
      van Merri{\"e}nboer, Bart  and
      Gulcehre, Caglar  and
      Bahdanau, Dzmitry  and
      Bougares, Fethi  and
      Schwenk, Holger  and
      Bengio, Yoshua},
    editor = "Moschitti, Alessandro  and
      Pang, Bo  and
      Daelemans, Walter",
    booktitle = "Proceedings of the 2014 Conference on Empirical Methods in Natural Language Processing ({EMNLP})",
    month = oct,
    year = "2014",
    address = "Doha, Qatar",
    publisher = "Association for Computational Linguistics",
    doi = "10.3115/v1/D14-1179",
    pages = "1724--1734"
}

@article{hochreiter1997lstm,
  title={Long short-term memory},
  author={Hochreiter, Sepp and Schmidhuber, J{\"u}rgen},
  journal={Neural computation},
  volume={9},
  number={8},
  pages={1735--1780},
  year={1997},
  publisher={MIT Press}
}

@inproceedings{lundberg2017unified,
  title={A Unified Approach to Interpreting Model Predictions},
  author={Lundberg, Scott M and Lee, Su-In},
  booktitle={Advances in Neural Information Processing Systems},
  volume={30},
  pages={4765--4774},
  year={2017},
  publisher={Curran Associates, Inc.}
}

@article{baryakova2023overcoming,
  title={Overcoming barriers to patient adherence: the case for developing innovative drug delivery systems},
  author={Baryakova, Tsvetelina H and Pogostin, Brett H and Langer, Robert and McHugh, Kevin J},
  journal={Nature Reviews Drug Discovery},
  volume={22},
  number={5},
  pages={387--409},
  year={2023},
  publisher={Nature Publishing Group UK London}
}

@article{park2019injectable,
  title={Injectable, long-acting PLGA formulations: Analyzing PLGA and understanding microparticle formation},
  author={Park, Kinam and Skidmore, Sarah and Hadar, Justin and Garner, John and Park, Haesun and Otte, Andrew and Soh, Bong Kwan and Yoon, Gwangheum and Yu, Dijia and Yun, Yeonhee and others},
  journal={J. Control. Release},
  volume={304},
  number={10.1016},
  year={2019}
}

@Inbook{Witika2025,
author="Witika, Bwalya A.
and Poka, Madan S.
and Makoni, Pedzisai A.
and Ramazani, Farshad
and Nkanga, Christian I.",
editor="Ramazani, Farshad",
title="Clinical Applications of Long Acting Injectables: Systemic and Local Drug Delivery",
bookTitle="Biodegradable Long Acting Injectables and Implants: An Industry View",
year="2025",
publisher="Springer Nature Switzerland",
address="Cham",
pages="185--220",
isbn="978-3-031-72302-5",
doi="10.1007/978-3-031-72302-5_7",
url="https://doi.org/10.1007/978-3-031-72302-5_7"
}

@article{Li03042015,
author = {Yang Li and Li Yang},
title = {Driving forces for drug loading in drug carriers},
journal = {Journal of Microencapsulation},
volume = {32},
number = {3},
pages = {255--272},
year = {2015},
publisher = {Taylor \& Francis},
doi = {10.3109/02652048.2015.1010459}
}

@article{Feng2015EffectsOD,
  title={Effects of drug and polymer molecular weight on drug release from PLGA‐mPEG microspheres},
  author={Shuibin Feng and Lei Nie and Peng Zou and Jinping Suo},
  journal={Journal of Applied Polymer Science},
  year={2015},
  volume={132},
  url={https://api.semanticscholar.org/CorpusID:86245356}
}

@article{Bickerton2012,
title = {Interconnected Roles of Scaffold Hydrophobicity, Drug Loading, and Encapsulation Stability in Polymeric Nanocarriers},
author = {Bickerton, Sean and Jiwpanich, Siriporn and Thayumanavan, S.},
year = {2012},
journal = {Molecular Pharmaceutics},
volume = {9},
number = {12},
pages = {3569--3578},
doi = {10.1021/mp3004226}
}

@article{klose2006porosity,
  title={How porosity and size affect the drug release mechanisms from PLGA-based microparticles},
  author={Klose, D and Siepmann, F and Elkharraz, K and Krenzlin, S and Siepmann, J},
  journal={International Journal of Pharmaceutics},
  volume={314},
  number={2},
  pages={198--206},
  year={2006},
  publisher={Elsevier},
  doi={10.1016/j.ijpharm.2005.07.031}
}

@article{siepmann2004effect,
  title={Effect of the size of biodegradable microparticles on drug release: experiment and theory},
  author={Siepmann, J and Faisant, N and Akiki, J and Richard, J and Benoit, J P},
  journal={Journal of Controlled Release},
  volume={96},
  number={1},
  pages={123--134},
  year={2004},
  publisher={Elsevier},
  doi={10.1016/j.jconrel.2004.01.011}
}

@article{LUAN2006168,
title = {Key parameters affecting the initial release (burst) and encapsulation efficiency of peptide-containing poly(lactide-co-glycolide) microparticles},
journal = {International Journal of Pharmaceutics},
volume = {324},
number = {2},
pages = {168-175},
year = {2006},
issn = {0378-5173},
doi = {https://doi.org/10.1016/j.ijpharm.2006.06.004},
url = {https://www.sciencedirect.com/science/article/pii/S0378517306004546},
author = {Xiaosong Luan and Marc Skupin and Jürgen Siepmann and Roland Bodmeier},
}

@article{Kim1992EffectOL,
  title={Effect of Loading on Swelling-Controlled Drug Release from Hydrophobic Polyelectrolyte Gel Beads},
  author={Cherng-ju Kim and Ping I. Lee},
  journal={Pharmaceutical Research},
  year={1992},
  volume={9},
  pages={1268-1274},
  url={https://api.semanticscholar.org/CorpusID:36790901}
}

@article{Mao2007EffectOW,
  title={Effect of WOW process parameters on morphology and burst release of FITC-dextran loaded PLGA microspheres.},
  author={Shirui Mao and Jing Xu and Cuifang Cai and Oliver Germershaus and Andreas K. Schaper and Thomas Prof Dr Kissel},
  journal={International journal of pharmaceutics},
  year={2007},
  volume={334 1-2},
  pages={
          137-48
        },
  url={https://api.semanticscholar.org/CorpusID:39902105}
}

@article{alhnan2010inhibiting,
  author    = {Alhnan, Mohamed A. and Cosi, David and Murdan, Sudaxshina and Basit, Abdul W.},
  title     = {Inhibiting the gastric burst release of drugs from enteric microparticles: the influence of drug molecular mass and solubility},
  journal   = {Journal of Pharmaceutical Sciences},
  year      = {2010},
  volume    = {99},
  number    = {11},
  pages     = {4576--4583},
  month     = nov,
  doi       = {10.1002/jps.22174},
  pmid      = {20845456}
}

@article{Park2025PLGA,
  author    = {Park, Kinam},
  title     = {PLGA-based long-acting injectable (LAI) formulations},
  journal   = {Journal of Controlled Release},
  volume    = {382},
  pages     = {113758},
  year      = {2025},
  month     = jun,
  doi       = {10.1016/j.jconrel.2025.113758},
  pmid      = {40268201},
  pmcid     = {PMC12065662},
  note      = {Epub 2025 Apr 21}
}

@article{Mishra_Waghamare_Khanage_2025, place={India}, title={Acarbose-Loaded PLGA Microspheres: Efficient Encapsulation and Controlled Release}, volume={15}, url={https://jddtonline.info/index.php/jddt/article/view/7165}, DOI={10.22270/jddt.v15i6.7165},  number={6}, journal={Journal of Drug Delivery and Therapeutics}, author={Mishra , Anurag and Waghamare , Suresh and Khanage, S. G.}, year={2025}, month={Jun.}, pages={30–40} }

\end{document}